\documentclass[11pt]{article}
\usepackage{graphicx}
\usepackage[margin=1.0in]{geometry}
\usepackage[usenames,dvipsnames]{color}
\usepackage{cite}
\usepackage{url}
\usepackage{authblk}
\usepackage{phonetic}
\usepackage{setspace}
\usepackage[colorlinks = true,
            linkcolor = blue,
            urlcolor  = blue,
            citecolor = blue,
            anchorcolor = blue]{hyperref}

\newcommand\pubnumber{Snowmass 2021 – Topical Group AF7RF}
\newcommand\pubdate{March 2022}

\def\Title#1{\begin{center} {\LARGE #1 } \end{center}}
\def\Author#1{\begin{center}{ \sc #1} \end{center}}
\def\Address#1{\begin{center}{ \it #1} \end{center}}

\newcommand\pubblock{\rightline{\begin{tabular}{l} \pubnumber\\
         \pubdate \end{tabular}}}
\newenvironment{Executive Summary}{\begin{quotation} \begin{center}
                       Executive Summary
     \end{center}\bigskip  }{\end{quotation}}

\def\beq{\begin{equation}}
\def\eeq#1{\label{#1}\end{equation}}
\def\eeqn{\end{equation}}

\newenvironment{Eqnarray}%
   {\arraycolsep 0.14em\begin{eqnarray}}{\end{eqnarray}}
\def\beqa{\begin{Eqnarray}}
\def\eeqa#1{\label{#1}\end{Eqnarray}}
\def\eeqan{\end{Eqnarray}}

\let\bar=\overbar

\def\lsim{\mathrel{\raise.3ex\hbox{$<$\kern-.75em\lower1ex\hbox{$\sim$}}}}
\def\gsim{\mathrel{\raise.3ex\hbox{$>$\kern-.75em\lower1ex\hbox{$\sim$}}}}

\def\del{\partial}
\def\Dslash{\not{\hbox{\kern-4pt $D$}}}
\def\dslash{\not{\hbox{\kern-2pt $\del$}}}
\def\pslash{\not{\hbox{\kern-2pt $p$}}}
\def\ETmiss{\not{\hbox{\kern-4pt $E$}}_T}

\def\Dlr{\mathrel{\raise1.5ex\hbox{$\leftrightarrow$\kern-1em\lower1.5ex\hbox{$D$}}}}

\def\MSB{{\bar{M \kern -2pt S}}}
\def\msb{{\bar{\scriptsize M \kern -1pt S}}}

\def\drb{{\bar{\scriptsize D \kern -1pt R}}}

\newcommand\snowmass{\begin{center}\rule[-0.2in]{\hsize}{0.01in}\\\rule{\hsize}{0.01in}\\
\vskip 0.1in Submitted to the  Proceedings of the US Community Study\\ 
on the Future of Particle Physics (Snowmass 2021)\\ 
\rule{\hsize}{0.01in}\\\rule[+0.2in]{\hsize}{0.01in} \end{center}}

\begin{document}

\pubblock

\Title{Next-Generation Superconducting RF Technology based on Advanced Thin Film Technologies and Innovative Materials for Accelerator Enhanced Performance \& Energy Reach}

\bigskip

\Author{A.- M. Valente-Feliciano{1}  \thanks{valente@jlab.org}, C. Antoine \uplett{2}, S. Anlage \uplett{9}, G. Ciovati \uplett{1}, J. Delayen \uplett{1, 11}, F. Gerigk\uplett{3}, A. Gurevich\uplett{11}, T. Junginger\uplett{15, 16}, S. Keckert\uplett{6}, G. Keppel\uplett{7}, J. Knobloch\uplett{6, 12}, T. Kubo\uplett{8}, O. Kugeler \uplett{5}, D. Manos\uplett{17},C.  Pira\uplett{7}, T. Proslier\uplett{2}, U. Pudasaini\uplett{1}, C.E. Reece\uplett{1}, R.A. Rimmer\uplett{1}, G.J. Rosaz\uplett{3},T. Saeki\uplett{8}, R. Vaglio\uplett{10}, R. Valizadeh\uplett{14}, H. Vennekate \uplett{1}, W. Venturini Delsolaro\uplett{3}, M. Vogel \uplett{12}, P. B. Welander\uplett{13}, M. Wenskat\uplett{4, 5}}

\medskip
\begin{footnotesize}

\Address{{\uplett1}Thomas Jefferson National Accelerator Facility, Newport News, VA 23602 USA}
\Address{\uplett{2}CEA Saclay DRF/IRFU/DACM, 91191 Gif-sur-Yvette France}
\Address{\uplett{3}European Organization for Nuclear Research (CERN), CH-1211 Geneva 23 Switzerland}
\Address{{\uplett4}Deutsches Elektronen-Synchrotron, 22607 Hamburg, Germany}
\Address{\uplett{5}Institute of Experimental Physics, University of Hamburg, 22761 Hamburg, Germany}
\Address{\uplett{6}Helmholtz-Zentrum Berlin, Albert-Einstein-Str. 15, 12489 Berlin, Germany}
\Address{\uplett{7}Istituto Nazionale di Fisica Nucleare, Legnaro National Labs (INFN-LNL), 35020 Legnaro PD, Italy}
\Address{\uplett{8}High Energy Accelerator Research Organization (KEK), Tsukuba, Ibaraki 305-0801, Japan}
\Address{\uplett{9}University of Maryland, College Park, MD 20742 USA}
\Address{\uplett{10}University of Napoli Federico II, 80125 Napoli Italy}
\Address{\uplett{11}Old Dominion University, Norfolk, VA 23529 USA}
\Address{\uplett{12}Universität Siegen, Walter-Flex-Str. 3, 57068 Siegen, Germany}
\Address{\uplett{13}SLAC National Accelerator Laboratory, Menlo Park, CA 94025, USA}
\Address{\uplett{14}STFC Accelerator Science and Technology Centre (ASTeC), Warrington, WA4 4AD United Kingdom}
\Address{\uplett{15}TRIUMF, Vancouver BC V6T 2A3 Canada}
\Address{\uplett{16}University of Victoria, Victoria BC V8P 5C2 CANADA}
\Address{\uplett{17}College of William and Mary, Williamsburg, VA 23185 USA}

\end{footnotesize}

\medskip
\snowmass
\medskip
 \begin{Executive Summary}
\noindent Superconducting radio frequency (SRF) systems, essentially based on bulk niobium (Nb), are the workhorse of most particle accelerators and achieve high levels of performance and reliability. Today’s exceptional performance of bulk Nb SRF cavities is the fruit of more than five decades of intensive research, essentially aimed at optimizing the material for thermal stabilization of defects, and significant funding efforts. Last incremental advances with several novel surface treatments are allowing Nb cavities to exceed previous record $Q$ factors and avoiding degradation with increasing gradients. These include nitrogen infusion and doping, oxygen alloying with a two-step baking process. The next generation of particle accelerators will require operational parameters beyond the state-of-the-art Nb capabilities. Superconducting RF has then to rely on superior materials and advanced surfaces beyond bulk Nb for a leap in performance and efficiency.
The SRF thin film development strategy aims at transforming the current SRF technology into a technology using highly functional materials, where all the necessary functions are addressed. The community is deploying efforts in three research thrusts to develop the next-generation thin-film based cavities. The first line of developments aims at investigating Nb/Cu coated cavities that perform as good as or better than bulk niobium at reduced cost and with better thermal stability. Recent results with greatly improved accelerating field and dramatically reduced Q-slope show the potential of this technology for many applications. In particular high current storage ring colliders such as FCC, EIC and CEPC, where the frequency is typically lower and the gradients are modest, could benefit greatly from the cost savings and operational advantages of this technology. The second research thrust is to develop cavities coated with materials that can operate at higher temperatures or sustain higher fields. Proof-of-principle has been established for the merit of Nb$_{3}$Sn for SRF applications. Research is now needed to further exploit the material and reach its full potential with novel deposition techniques. The third line of research is to push SRF performance beyond the capabilities of the superconductors alone with multi-layer coatings. In parallel, developments are needed to provide quality substrates, innovative cooling schemes and cryomodule design tailored to SRF thin film cavities. 

Recent results in these three research thrusts suggest that SRF thin film technologies are at the eve of a technological revolution. However, in order for them to mature, active community support and sustained funding are needed and should address fundamental developments supporting material deposition techniques, surface and RF research, and technical challenges associated with scaling, automation and industrialization.
 With dedicated and sustained investment in SRF thin film R\&D, next-generation thin-film based cavities will become a reality with high performance and efficiency,  facilitating energy sustainable science while enabling higher luminosity, and higher energy.

\end{Executive Summary}

\def\thefootnote{\fnsymbol{footnote}}
\setcounter{footnote}{0}

\section{Introduction}
 Superconducting RF Technology has been the building block for many high-energy physics accelerators, essentially based on bulk niobium (Nb). Over the pastn five decades, the RF performance of bulk Nb cavities has continuously improved with incremental material and surface developments \cite{Padamsee2020}. With nitrogen surface doping \cite{Bafia 2019} as of late, 1.3 GHz cavities reliably achieve $Q_{0}$ around $4 x 10^{10}$ at 35 MV/m and 2 K. Nibbling at last improvements in cavity performance to reproducibly reach the intrinsic limits of niobium is becoming increasingly difficult and exponentially expensive. Long-term solutions for superconducting radio-frequency (SRF) surfaces efficiency enhancement need to be pursued and require a technological leap to produce the next generation of SRF cavities with cost-effective and reliable production methods scalable to mass industrialization.  The greatest potential for dramatic new performance capabilities lies with methods and materials, which deliberately produce the sub-micron thick critical surface layer in a controlled way. The development of SRF surfaces with thin films of Nb and other superconducting materials has the potential to improve further SRF cavity performance beyond the Nb sheet material intrinsic limits. This development combined with higher efficiency and quality control is most relevant for next-generation structures required for future SRF accelerators. The potential of fully engineered,functional, and truly application-tailored SRF surface strategies would dramatically reduce capital and operating costs for future accelerators, lowering the cost-of-entry of SRF technology, both for large scale facilities and compact accelerators. Although it is beyond the focus of this White Paper, SRF thin films produced by Advanced coing techniques have also upcoming applications beyond the realm of accelerators, in superconducting electronics,sensing, and quantum information systems.

\section{Next Generation SRF Cavities Based on Advanced Coating Technology for CW Accelerators}

Electron-ion colliders, electron coolers, energy recovery linacs (ERL), X-ray light sources and free electron lasers (FELs) require continuous wave (CW) operation. Minimal cavity residual losses and minimized cryogenic consumption, i.e. high quality factors, $Q$, at medium gradients are then more essential than high gradients. Furthermore, reducing overall power consumption is of primary importance for high current CW facilities thus pursuing superconducting acceleration at liquid He (LHe) temperature, 4.2 K present a considerable advantage and is a must.  

CW electron accelerators find other high energy physics applications such as in dark matter research \cite{Achenbach-LOI76,Nosochkov-IPAC17} and high duty factor linear accelerators (proposed driver for future multi-MW neutrino experiments at Fermilab for LBNF/DUNE \cite{Abi2020} as well as a driver for a muon collider).  

Historically, superconducting accelerators for HEP in the United States have been based on bulk Nb technology. However, over the years, niobium film on copper (Nb/Cu) technology has positioned itself as an alternative route for the future of superconducting structures used in accelerators. Superconducting Nb/Cu cavities produced by the magnetron sputtering technology have been successfully deployed in Europe at CERN with LEP-II II (352 MHz, 4.2 K) and LHC (400 MHz, 4.2 K). This technology is also employed in other accelerator facilities, such as ALPI (INFN -LNL \cite{Sladen 1998}) or HIE-ISOLDE (CERN). It is the technology of choice for the proposed circular electron positron $e^{+}e^{-}$) collider FCC-ee \cite{FCC-ee}, which deploys hundreds of multi-cell SRF cavities at 400 MHz for a total RF voltage of multiple GeV at 4.2 K. The subsequent $t\bar{t}$ machine foresees operation above 10 GeV with a combination of 400 and 800 MHz multi-cell cavities\cite{FCC-hh}.

However, although exhibiting very high $Q$ at low field, Nb/Cu cavities have been plagued with steep RF losses ($Q$-slope) at medium to high gradients. Recently, R\&D efforts, jointly at JLab and CERN, to develop advanced coating technologies (energetic condensation) have produced encouraging results alluding to the promise of superior performing SRF films \cite{Valente-Feliciano 2015, Rosaz 2015, Burton 2019, Avino 2019, CERN 2021}. These techniques, along with careful engineering of surfaces \cite{Delsolaro 2019} and Nb thick film approach at LNL \cite{Palmieri 2017}, have produced Nb/Cu films with bulk-like material properties, greatly improved accelerating field and dramatically reduced $Q$-slope, demonstrating that the early sputtered Nb/Cu cavity performance is not a fundamental limitation. Further effort is needed to scale up these processes and develop protocols and procedures that can be industrialized. 

The past decade has seen, regain interest for the development of Nb$_{3}$Sn on niobium. With its critical temperature of about 18 K, the A15 Nb$_{3}$Sn can achieve high $Q_{0}>10^{10}$ at 4.2 K . Maintaining intrinsic quality factors in the $10^{10}$ to $10^{11}$ range would reduce energy consumption and thus cryogenic operating costs by as much as an order of magnitude and would substantially decrease infrastructure costs for the cryogenic plant. It has been demonstrated that Nb$_{3}$Sn coated single-cell cavities can reach $> 18 MV/m$ with $Q$ above $10^{10}$ at 4.2 K.

Successful SRF thin film cavity development will lead to significant savings in cavities and cryomodule costs. It is in line with the demands of the future generation of particle physics machines such as the electron – ion collider (EIC), FCC-ee \cite{FCC-ee,Aull 2016} which require R\&D pushing the state of the art and enabling technological breakthroughs to provide the required reliability and keep costs within acceptable levels. Accelerators relying on operation at 4 K and/or on low frequency (large) cavities for which the cost of bulk Nb material becomes prohibitive (precision muon generation) or high-power compact proton and CW ion linear accelerators for isotope production facilities (FRIB) would also benefit.

\section{Higher Gradients \& High Q with Advanced Materials \& Structures}

To fulfill the need for ever higher particle energies for HEP experiments, increasing the accelerating gradient is an absolute necessity so facilities are kept to a reasonable size. Another primary objective is to frame both construction and operating costs. Thus, making progress towards more affordable accelerating RF systems at an industrial scale is also mandatory. Engineering programs aiming at optimizing the fabrication cost of such systems will then have increased priority in the coming years. 

Some key factors may see their importance growing with the size of HEP facilities or with the required beam parameters becoming more and more difficult to achieve. Energy efficiency is definitively a key parameter for future HEP facilities, which will likely need to limit their electricity consumption. 

Alternative materials with higher critical temperature and critical field are prime candidates to surpass the established bulk Nb technology \cite{Valente-Feliciano 2016}. Materials such as Nb$_{3}$Sn offer order of magnitude improvements in operating efficiency, and a theoretical pathway to 100 MV/m gradient \cite{Rimmer LOI SNOWMASS 2021}. Recent R\&D efforts \cite{Posen 2019, Eremeev 2020, Pudasaini 2019, Hall 2017} have demonstrated that the persistent Q-slope and gradient limitation observed in the past \cite{Kneisel 1996} are not fundamental but process induced and therefore amenable to improvement. Alternative deposition approaches such as sputtering, energetic condensation and atomic layer deposition (ALD) should be fully explored for enhanced properties and conformality. Early results with sequential and stoichiometric deposition both on Nb and on Cu are promising \cite{Sayeed 2018, Rosaz 2015b} and could prove to push the Nb$_{3}$Sn technology further. Other materials such as NbTiN, NbN, V$_{3}$Si, etc. should also be re-evaluated with advanced coating techniques. Newly discovered high temperature superconducting (HTS) materials (pnictides \cite{Si 2016} …) would be particularly interesting if any of them turns out to have favorable microwave properties \cite{Valente-Feliciano 2016}. 

Pushing the performance limits further, the combination of such materials with adequate dielectric material in multi-layered structures have been conceived as a performance enhancer for bulk Nb and Nb/Cu film cavities. Theoretical models \cite{Gurevich 2015, Kubo 2021} predict that appropriately fabricated nanometric superconductor-insulator-superconductor (SIS) multilayer films can delay vortex penetration in Nb surfaces allowing them to sustain higher surface fields than any pure material.

\section{Compact SRF Accelerators for Societal Applications}

The vast majority of particle accelerators in operation today are for industrial and medical applications. Many of them operate in continuous wave (CW) mode, all of them are based on conventional i.e. normal conducting technology. As SRF accelerator technology has replaced the normal conducting approach for high power machines in science due to its dramatically superior efficiency, it offers the same advantage for such applications. Even beyond that, the ability to generate high power CW beams enables completely new applications for industry and societal applications. Examples include environmental remediation (water and waste treatment/ sterilization); cancer therapy, sterilization of medical instruments, medical isotopes production; radiation crosslinking of plastics and rubbers; engineering of material and surfaces with altered properties; radiation driven chemistry; food preservation. Other applications include accelerator-driven systems (ADS) for nuclear waste transmutation or power generation, and high-intensity proton accelerators for homeland security (nuclear weapons and materials detection).

These applications need a new class of compact, mobile, high-power electron accelerators, small and light enough to be located on a mobile platform or in a university or hospital room and with a drastically reduced cryogenic refrigeration requirement.  For such compact-sized SRF machines, energy efficiency is one of the most critical areas of development. The use of superconducting thin films, enabling high-Q operation of cavities at significantly higher temperatures, is key as they do not require the installation and operation of complex, subatmospheric cryoplants. 

First proof-of-principle experiments have demonstrated successful operation of Nb$_{3}$Sn coated cavities cooled with commercial, off-the-shelf cryocoolers, paving the way for completely cryogen-free operation for SRF based accelerator technology for  applications beyond science \cite{Dhuley 2020, Ciovati 2020, Stilin 2020}.  

The emergence of reliable, energy efficient high $Q$ systems, based on highly performing film-based SRF cavities along with transformative development with cryocoolers  would combine into cost effective compact superconducting accelerators with reduced footprint and capital investment, transformational for societal applications.

\section{SRF Thin Film Technology Research Thrusts}
\begin{figure}
	\begin{center}
		\includegraphics[width=1\hsize]{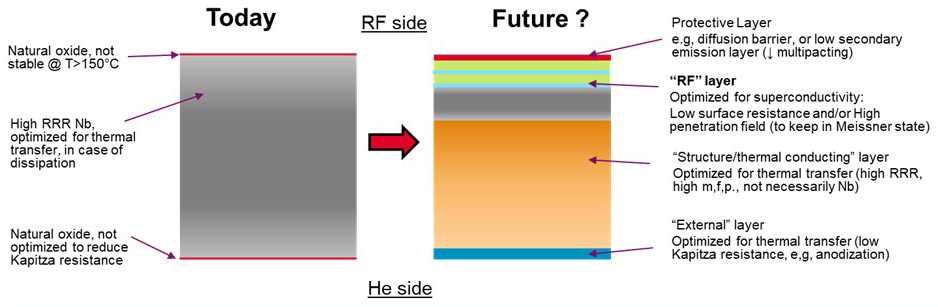}
	\end{center}
	\caption{Path toward more efficient functional SRF materials (courtesy C.Z. Antoine)}
	\label{fig:Picture 2}
\end{figure}

Due to the very shallow penetration depth of RF fields (only about 40 nm for Nb), SRF
properties are inherently a surface phenomenon, involving a material thickness of less
than 1 $\mu$m. 
Instead of using bulk Nb, there is then merits of depositing a superconducting film on the inner surface of a castable cavity structure made of copper (Cu) or
aluminum (Al). At the system design level, this would allow one to decouple the active
SRF surface from the accelerating support structure and its cooling, bringing dramatic
change to the cost framework of SRF accelerators.
Thin film technology provides then the framework for developing functional SRF surfaces and materials (see figure~\ref{fig:Picture 2} for illustration) that address the necessary functions: 

\begin{list}{-}{}
	\item Mechanical support structure and thermal conduction: high purity niobium, copper  
	\item Optimized superconducting layer: high quality Nb film, higher-$T_{c}$ and $H_{sh}$ material, SIS multilayers
	\item Capping layer (protection against atmosphere, reduction of multipacting…) 
	\item In a later step, other functionalization like e.g. an external layer optimized to reduce thermal interface resistance with Helium (Kapitza resistance) can be considered.
\end{list}

%%%%%%%%%%%%%%%%%%%%%%%%%%%%%%%%%%%%%%%%%%%%%%%%%%%%%%%%%%%%%%%%%%%%%%%%%
%%%%%%%%%%%%%%%%%%%%%%%%%%%%%%%%%%%%%%%%%%%%%%%%%%%%%%%%%%%%%%%%%%%%%%%%%%% 

The development of functional materials offers the opportunity to tailor the final performance of the SRF system to the machine specifications. For instance doping treatment optimize the superconducting surface of Nb by modifying the surface m.f.p. without affecting the highly thermally conductive pure Nb deeper in the material. It increases $Q_{0}$ at medium field but is detrimental at high field. One could progress further with more functional materials.
Combining the different aspects that are already under study, one has the potential to increase the $Q_{0}$ by a factor of 10 and the accelerating field by a factor of 2 or 3, for the materials currently explored. The reduction on both investment and operation costs of SRF machines would be tremendous. Several of these techniques (e.g. multilayers) can be used to upgrade existing facilities as they can be directly applied on bulk Nb. For instance depositing multilayers inside ILC cavities has the potential to upgrade the facility to its final energy without –in principle- the need to expand tunnels.

The community is working on three parallel research thrusts: Nb/Cu, alternate materials on Nb or Cu substrates, composite materials based on nanometric SIS multilayers.

\subsection{Nb/Cu Technolgy}

Nb films present a clear advantage compared to bulk Nb for defined accelerator parameter sets, in particular for low cavity frequencies and/or for operation at 4.2 K. The BCS surface resistance for thin films is lower than for bulk Nb, due to a normal state electrical resistivity close to the theoretical optimum.

Decoupling the SRF surface from the cryogenic system by depositing a thin layer of Nb on the inner surface of a Cu cavity offers several advantages:  

\begin{list}{-}{}
	\item Increased temperature stability due to the Cu substrate higher thermal conductivity 
	\item Operation at 4.2 K, generating capital and operational cost savings 
	\item Material cost saving, particularly for low frequency structures 
	\item Easily machinable and castable structures, opening perspectives for significant cryomodule simplification. 
\end{list} 

However, although exhibiting very high $Q$ at low field, Nb/Cu cavities have been plagued with steep RF losses ($Q$-slope) at medium to high gradients. Recently, R\&D efforts, jointly at JLab and CERN, to develop advanced coating technologies (energetic condensation) have produced encouraging results alluding to the promise of superior performing SRF films \cite{Valente-Feliciano 2015, Rosaz 2015, Burton 2019, Avino 2019, CERN 2021}. These techniques, along with careful engineering of surfaces \cite{Delsolaro 2019} and Nb thick film approach at LNL \cite{Palmieri 2017}, have produced Nb/Cu films with bulk-like material properties, greatly improved accelerating field and dramatically reduced Q-slope, demonstrating that the early sputtered Nb/Cu cavity performance is not a fundamental limitation. Further effort is needed to scale up these processes and develop protocols and procedures that can be industrialized.

\subsection{Alternate superconducting materials on Nb or Cu substrates}
Among the thousands superconductors available, only a limited number of compounds with a $T_{c}$ higher than Nb (9.27 K) have a true potential for SRF cavity applications. For a superconducting material to be a good candidate for SRF applications, it needs not only a high $T_{c}$ but also a low resistivity in the normal state to minimize RF losses. Additionally, high thermodynamic critical, $H_{c}$, superheating, $H_{sh}$, and lower critical, $H_{c1}$, fields are necessary to maximize accelerating gradients. Many compounds such as NbTiN, NbN, Nb$_{3}$Sn, MgB$_{2}$ have higher critical temperatures and critical magnetic fields than Nb.
 However, their $H_{c1}$ is generally lower than Nb and, at present, none of these materials match Nb in terms of its RF performance or ease of use for accelerator applications.
 Some of the compounds such as NbN and NbTiN, have been evaluated for SRF applications in de 80-90s. 
 
\subsubsection{Nb$_{3}$Sn and other A15 compounds}

The class of inter-metallic A15 materials  has been the object of interests for SRF applications for a certain number of years, due to their transition temperature typically a factor of 2 higher than Nb. The use of such materials for accelerating structures necessarily relies on the elaboration of thin films as the brittle nature of these compounds prevents their use as bulk material.

The most widely developed technique to form the A15 Nb$_{3}$Sn phase on Nb is Sn vapor diffusion in bulk Nb, initially developed by Siemens in the 90s and rekindled in 2010s by Cornell, Fermi Lab and Jefferson Lab. Improvements in understanding the materials science and refinement of the coating process have led to resolving the precipitous $Q$-slope, increasing the maximum attainable accelerating gradient by a factor of 5 with $Q_{0}>10^{10}$ at 4.2 K \cite{Muller 2000,Posen 2014,Posen 2015,Posen 2021}. However, Nb$_{3}$Sn cavity performance has yet to reach its full  potential.  

The applications of Nb$_{3}$Sn to SRF cavities faces several challenges particularly due to the brittle nature of the material which demands novel tuning methods for cavity operation. A detailed description of the status and objectives of the research on Nb$_{3}$Sn/Nb  is found in the dedicated White Paper submitted to Snowmass 2021 \cite{Nb3Sn-SnowmassWP}.

The recent progress in material understanding and RF performance of Nb$_{3}$Sn films on Nb has stimulated widened interest and alternative techniques are emerging, specifically for deposition in Cu cavities. The attraction is for A15 materials deposited onto Cu substrates is to improve the quality factor of coated Cu cavities by a factor of 10 at an operating temperature of 4.2 K. Alternate Nb$_{3}$Sn R\&D developments with novel deposition techniques have started at CERN, IMP, ULVAC/KEK, NHMFL/Florida State University/University of Texas–Arlington, Peking University, STFC, ODU, and Ultramet \cite{Ilyina 2019, Yang SRF2019,Ito SRF 2019,Ziao SRF19,Kim 2018, Valizadeh IPAC 2019, Sayeed SRF2019, Ge 2019}.

Beyond Nb$_{3}$Sn, other A15 materials of interests are V$_{3}$Si, Nb$_{3}$Al, and Mo$_{3}$Re. They have similar $T_{c}$s, some of these compound may be  easier to deposit with PVD. Thus far, the exploration of these materials has been marginal for SRF applications. 

\subsubsection{MgB$_{2}$ }

With a $T_{c}$ of 40 K, MgB$_{2}$ has raised interest since its discovery in 2001.
This compound has the particularity of having 2 superconducting gaps [124, 125] (with $\Delta_{p}=2.7 meV$ and $\Delta_{p}=6.7 meV$).  The RF response has shown lower energy gap behavior. This lower gap is still about twice the gap of Nb (1.5 meV). The normal resistivity can also be made quite low (best values are $\leq{}$1$\mu{}$$\Omega{}$cm). With a higher $T_{c }$ than Nb, a low resistivity, a larger gap, a higher critical field, there appears to be room for better performance than Nb and for operation beyond 4.2 K.
In the last few years,consistent efforts have led Temple University to develop high quality MgB$_{2}$ films with potential for SRF applications. In collaboration with ANL, they recently started to deploy this material on 3D structures at 3 GHz.
One of the challenges with using MgB$_{2}$ for SRF application is the degradation of the film properties with exposure to moisture (the surface resistance $R_{s}$ increases and $T_{c }$ degrades). One way to overcome such degradation during the common cavity surface cleaning procedures may be to use a cap layer to protect the MgB$_{2}$ film.
With potential operation at 20 K, MgB$_{2}$, especially when used in the form of multilayers, is attractive for SRF cavity operation in cryocoolers \cite{Tan 2016}.

\subsection{superconductor-insulator-superconductor (SIS) nanometric layers }
 
 Despite higher $T_{c}$ and the potential for a higher superheating field $H_{sh}$, most good SRF material candidates do not outperform the properties and ease of use of Nb. So, in pursuit of increased maximum operation gradients ($E_{max}$), a nanometric superconductor/insulator/superconductor (SIS) structure concept has been proposed for its potential capability of supporting otherwise inaccessible surface magnetic fields in SRF cavities \cite{AG2006}. The idea is to coat an SRF cavity (bulk Nb or thick Nb ﬁlm) with a nanometric multilayer structure (SIS structures) composed of alternating superconducting (S) and insulating (I) layers with a thickness $d_{S}$ smaller than the penetration depth $\lambda_S$. If the superconducting ﬁlm S is deposited with a thickness $d_{S}<<\lambda$, the Meissner state can be retained at magnetic ﬁeld much higher than bulk $H_{c1}$ as for $d_{S}<<\lambda$.
The strong increase of ﬁrst ﬂux penetration ﬁeld $H_{fp}$ in ﬁlms allows the surface to support RF magnetic ﬁelds higher than the lower critical ﬁeld $H_{c1}$ of bulk Nb (ﬁgure 2). The thin insulating layer I inhibits vortex oscillations and prevents global vortex penetration in the cavity. Its thickness, $d_{I}$, needs to be big enough to avoid Josephson junction effects. The BCS resistance is also strongly reduced because of the use of superconductors with a higher gap $\Delta$ (Nb$_{3}$Sn, NbTiN). The effect can be furthered by adding SI layer sequences. The optimum number of SI layers is determined by a compromise between vortex dissipation reduction, superconductivity suppression at the S–I interfaces, and thermal conductivity. With such optimized SIS structures, $Q$-values at 4.2 K two orders of magnitude above Nb values are predicted, and the sustainable external magnetic ﬁeld could be more than doubled. The multilayered approach composed of nanometric superconducting (50–200 nm) and insulating (5–10 nm) thin film stacks has the potential to significantly increase $E_{max}$ by 20 to 100 \% as compared to Nb. This solution can be applied to any optimized thin film mentioned in the previous section. The major challenge is to demonstrate  the feasibility of this solution for higher gradients i.e. $> 50 MV/m$. A 50 \% increase in the maximum accelerating gradient implies a construction cost saving for an XFEL-scale accelerator of about 100 M\$ and a 50 \% lower cryomodule operational cost.

Reﬁnements have been introduced \cite{AG15,Kubo17,AGKubo17} in order to accurately understand the ﬁeld limit in the multilayered SIS structure and taking into account surface roughness, and layer thickness non-uniformity, but further theoretical understanding are yet necessary. The technological challenges involved in adequate SIS multilayered structures include creation of high quality superconductors and dielectric materials with sharp, clean, transitions, and careful thickness control.

\section{Technology Development Paths}

Thin film based SRF technology is a multiplexed field and fast-tracked success requires a strategy with multiple areas of development to be pursued in parallel.

All three research thrusts presented in the previous section have commonalities in their respective required development paths: fundamental studies, advanced coating techniques tailored to the materials, quality substrates, SRF properties assessment, scalability studies. 

\subsection{Fundamental studies }

\subsubsection{Theoretical and material studies} 
To gain in-depth understanding of the fundamental limitations of thin film superconductors under radio-frequency fields.  Superconducting and RF properties are inherent to the material, so there is a clear need to develop a detailed understanding of the precise microscopic causal links between surface coatings/treatments and ultimate RF performance at cryogenic temperatures.  An understanding of these links would provide a clear roadmap for improvement of SRF cavity performance.  Materials studies include advanced material analysis techniques to characterize local defects and nanostructure, both laterally and as a function of depth, to characterize the electrodynamic properties of the near-surface region (where RF interacts) of superconductors.  Such studies, coupled with theory, can efficiently guide the development of new thin film materials and processes for SRF applications. 

\subsubsection{Theoretical studies for multilayered structures} 

Revealing the true SRF performance potential of SIS multilayers and optimizing their geometry requires addressing the following outstanding theoretical issues: 

\paragraph{Calculations of the dynamic superheating field and maximum accelerating gradient for the cavity SIS multilayer coatings.}  One of the important questions is how far the maximum breakdown field can be increased by tuning the SIS geometry. This question can be addressed by solving rather complicated equations of non-equilibrium superconductivity \cite{Kopnin01} to calculate the dynamic superheating field $H_{sh}(\omega, T, d)$ as function of frequency $\omega$, temperature $T$ and thicknesses $d$ of S and I layers which would give the optimum layer thicknesses which maximize $H_{sh}(\omega, T, d)$. It was shown recently that the dynamic superheating field $H_{sh}(T)$ near the transition temperature $T_{c}$ can exceed the static superheating field $H_{sh}(T)$ by factor 2 \cite{Sheikh20}.  

\paragraph{The effect of multilayer geometry on the surface resistance.} Because of the scattering of electrons on S-I interfaces, the surface resistance $R_{s}(l,d)$ in multilayers depends not only on the bulk mean free path $l$ but also on the layer thickness $d$ and the mechanism of interface scattering. Other factors which can strongly affect the residual surface resistance include broadening of the density of states by subgap states \cite{AGKubo17,KuboAG19,Kubo22} and blocking the RF currents by grain boundaries \cite{Sheikh17,Makita22}. Optimization of $R_{s}(l,d)$ in multilayers will require solving kinetic equations for the distribution function of quasiparticles in RF field \cite{Kopnin01} to find the values of $l$ and $d$ which minimize the surface resistance.        

\paragraph{High-field surface resistance and nonequilibrium superconductivity in multilayers.} Strong RF field in cavities causes oscillations in the density of states of quasiparticles and drive them out of equilibrium, making $R_{s}(l,d,H_{a})$ dependent on the field amplitude $H_{a}$ and changing the frequency dependence of $R_{s}(l,d,H_{a})$ as compared to the low-field Mattis-Bardeen theory. These mechanisms, as well as two-level atomic systems likely associated with amorphous niobium oxide at the surface \citen{Rom17}, can produce the $Q$ rise \cite{AG14} observed on alloyed Nb cavities. Theoretical account of these phenomena requires solving kinetic equations for quasi-particles along with temporal oscillation of superfluid density, superconducting gap and caused by the RF current pair breaking \cite{AG14,AG17, Kubo17,Kubo22}. The outcome of this challenging project is the calculation of the field-dependent quality factor and determination of the conditions under which the extended $Q(H)$ rise can reverse the medium and high-field $Q$ slope up to the breakdown field. This theory can be applicable to new SRF materials like Nb$_{3}$Sn \cite{Sundahl21} and pnictides, as well as to alloyed Nb multilayers. 

\paragraph{Reduction of the RF dissipation caused by penetration of vortices is one of the key advantages of SIS multilayers.} Small vortex semi-loops can penetrate from materials surface defects and then be intercepted by I layers, which stop further propagation of vortices in the bulk \cite{AG15,AG17, Kubo17, Liarte17}. The power dissipated by vortices in this process is an important characteristic that quantifies a tolerable level of vortex losses which do not appreciably deteriorate the cavity performance and does not trigger thermal runaway. The vortex losses can be calculated by numerical simulation of the time-dependent Ginzburg-Landau equations. 

Addressing these issues would allow to establish the fundamental theoretical performance limits of multilayers to raise the maximum accelerating gradient and quality factors by using SRF materials other than Nb. Another opportunity would be to use N-doped "dirty" Nb multilayers to reverse the reduction of the breakdown field and the high-field $Q$ slope, producing an extended $Q$-rise effect in the field range exceeding the superheating field of Nb \cite{AG15,AG17, Kubo17,Kubo22}. Here the theory would guide the ongoing materials optimization of thin film coated SRF cavities.  

\subsection{Advanced coating technologies and SRF film structure developments}

In praticality, achieving fully engineered functional SRF thin film material involves: 

\begin{list}{-}{}
	\item Developing and mastering advanced thin film deposition techniques in terms of final composition and structure. 
	\item Transferring known deposition techniques to the internal complex shape of the cavities. 
	\item Mastering quality of interfaces (substrate preparation, interlayers). 
	\item Establishing the proper compromise between optimum superconducting quality and fabrication cost (choice of the superconducting material). 
\end{list}

\subsubsection{Energetic condensation} 

The persistently recurrent $Q$-slope for Nb/Cu cavities has been shown to be partially due to porosities present in the films elaborated by Direct Current Magnetron Sputtering (DCMS). This issue is related to the low ionization ratio of the sputtered material (about 3\%) combined to a low energy distribution function of the sputtered species. The use of energetic condensation techniques such as Electron Cyclotron Resonance (ECR) \cite{Valente-Feliciano SRF2015} or High Power Impulse Magnetron Sputtering (HiPIMS) \cite{Valente-Feliciano SRF2013, Rosaz 2015} has proven to be the path to be used in order to elaborate dense films with improved characteristics thanks to the combination of: 

\begin{list}{-}{ }
	\item High ionization ratio of the deposit material ($>70 \%$) 
	\item Higher ion energy
	\item Energy control of the film species with substrate biasing. 
Recent development of positive short kick (40 $\mu$s) in HIPIMS \cite{Avino 2019} further simplify the target-substrate arrangement.
\end{list} 

These techniques are characterized by a number of surface and sub-surface processes that are activated or enabled by the added energy of the ions arriving at the surface \cite{Colligon 1995, Sarakinos 2010, Monteiro 2001, Anders 2010}. Among them are desorption of residual gases from the substrate surface, chemical bonds breaking, increased surface mobility, sub-implantation, creation and annihilation of defects. Such competing processes induce changes in film growth 
affecting nucleation processes, film adhesion, morphology, microstructure  and intrinsic stress.  
This allows approaching film growth in sequential steps : engineered interface (intermixing, interlayer), nucleation, adaptive template, subsequent growth and final RF surface. Each step can be energetically optimized or tuned depending on the substrate nature, the film material, or the desired film structure and properties (density, crystal orientation, low-temperature epitaxy...). The conformality of deposition is also greatly improved compared to DCMS and other conventional PVD techniques.

Recent results using both techniques have proven that the $Q$-slope problem can be efficiently addressed and resolved \cite{Burton 2019, CERN 2021}. A strong effort is now required to push the performance of the 1.3 GHz test cavities with the aim of reaching quality factor of $10^{10}$ at 20 MV/m of accelerating gradients. Such performance would place the Nb/Cu technology as a strong competitor to bulk Nb.  

Energetic condensation techniques can also benefit alternative SRF  material developments for cavity applications by lowering the deposition temperature necessary to form the superconducting phase of choice (A15 for Nb$_{3}$Sn, $\delta$-phase for Nb B1-compounds) thus increasing compatibility with low-temperature melting substrates such as Cu or Al. 

\subsubsection{Atomic Layer Deposition (ALD)} 

Achieving conformal deposition is a prime objective for complex SRF cavity geometries, particularly when considering composite materials such as SIS structures. Within the last decade, Atomic layer deposition (ALD) has developed from a pure R\&D technology to industry applications. Specifically, the semiconductor and the solar cell industry have implemented ALD in their production lines. ALD belongs to the class of chemical vapor deposition methods. It is inherently a self-limiting process with near perfect thickness control at the atomic level, and thus does not require a line of sight to the substrate. In principle, complex structures can then be coated without the difficulties typically encountered with sputtering, albeit the coating rates are very low. Preliminary demonstration has been provided  with the deposition in 1.3 GHz cavities of insulating layers synthesized via ALD \cite{Kalboussi2021}. As the surface chemistry and not the absolute mass flux of deposition species determines the film growth, ALD presents an undeniable advantage over all other physical vapor deposition methods with respect to coating conformality on nano- and micro-structured substrates because. 

Unfortunately, ALD synthetization of metallic superconducting materials is proven difficult and is still a niche field. Thus far, ALD is not compatible with state-of-the-art Nb or Nb$_{3}$Sn films due to the need of precursors typically based on elements incompatible with preserving the material purity required for high performance thin films. However, ALD deposition of superconducting materials such as NbN, NbTiN, MoN \cite{CCao2014} and MgB$_{2}$ \cite{ANL-MgB2} is making good progress. Its long-term potential for high performance 4.2 K (and above) systems may yet be greater than that of both the vapor-infusion and sputtering techniques. 

As a unique tool for surface engineering, ALD can find other applications in particle accelerators besides the synthesis of superconducting alloys and multilayers as the main SRF material:

\begin{list}{-}{}
	\item Secondary electron yield mitigation by the controlled deposition of very thin films (~ 1 nm) on RF exposed surfaces in vacuum \cite{Kalboussi2021}. 
	
	\item ALD can also be used to grow diffusion barriers to study the role of various alloys (Oxides, nitrides) on SRF cavity performances.  
 
	\item ALD deposited thin films with post annealing treatment can serve as dopant sources for Nb technology.  

	\item Adhesion and passivation layers deposited on copper by ALD can help maintaining long term stability in air exposed Cu surfaces and improving superconducting films adhesion reproducibility.  
 
	\item Difference between substrate and film thermal expansion could also be addressed by ALD engineering an intermediate layer with tuned thermal expansion coefficient.
\end{list} 

\subsubsection{Hybrid deposition techniques} 

Superconducting alloy materials such as NbN, NbTiN and some A15 compounds (Nb$_{3}$Sn, Nb$_{3}$Ge, V$_{3}$Si, Mo$_{3}$Re)  have been successfully deposited on various flat substrates either directly with sputtering of a stoichiometric target or by co-sputtering (silmutaneous sputtering of two constituents). In co-sputtering, the achieved composition is dependent on the relative positions of the target and the substrate. The perfect stoichiometry can then be obtained by manipulating these positions. However, the control of the stoichiometry may be difﬁcult over the large areas of accelerating cavities, especially if the stoichiometry range for the B1 or A15 phase is narrow. In this case, combining cylindrical sputtering with plasma enhancement by ECR allows to circonvent some of these issues and improve on deposition conformality.

Chemical vapor deposition (CVD) and plasma enhanced CVD (PECVD) of either single (Nb) or alloy superconducting material (NbTiN and NbN) has also been used to deposit mainly on flat substrates. In this process, one or more precursors, present in vapor phase, chemically react and form a solid ﬁlm on a substrate at the appropriate temperature. The deposition rate and the structure of the ﬁlm depend on the temperature and the reagent concentration. The control of the temperature and gas ﬂow uniformity over the entire cavity surface may be difﬁcult with complex geometries. As for ALD, the use of precursor may affect the purity and final properties of the films deposited. 

Combination of the two-deposition process of PVD and CVD can overcome the shortfall of each individual process. Extensivley developed at Temple University, the hybrid physical chemical vapor deposition (HPCVD) which combines physical and CVD has been shown to produce high quality MgB$_{2}$ thin films \cite{Wolak 2014,Wenura 2017}. The high temperature used in HPCVD favors excellent epitaxy and crystallinity, yielding RRR values in excess of 80. This technique is now being deployed on 3D structures such as a 3 GHz single-cell cavity \cite{Guo 2020}. another two-step process is being developed at ANL with a B layer deposited by CVD and post-reacted with Mg vapor to form MgB$_{2}$ \cite{Tajima 2015}.

Recently, the HPCVD process was modified to synthesize thin films of superconducting alloys on 3D geometry substrates by using magnetron sputtering with a single element target as the source for one of the elements of the compound and providing the other element in vapor form through a precursor. The plasma generated by magnetron sputtering facilitates the decomposition of the precursor and hence allows the needed chemical reaction to take place at a much lower temperature, compatible with a copper cavity. 

In the pursuit of inexpensive and low temperature methods to form the A15 phase of Nb$_{3}$Sn, another hybrid technique of interest as of late is Nb/CuSn with post diffusion  \cite{AltNb3Sn-SnowmassWP}, based on the bronze method. The principle is to deposit a Nb thin film on a bronze (CuSn) substrate and post-anneal it to favor the diffusion of Sn into the Nb and form the A15 phase. One of the major attraction of the method is that CuSn can be easily and inexpensively cast into a wide range of geometries. A dedicated White Paper has been submitted to Snowmass 21 \cite{AltNb3Sn-SnowmassWP}.

\subsubsection{Development of superconductor-insulator-superconductor (SIS) nanometric layers  and associated cavity deposition methods to further enhance the performance of bulk Nb and Nb/Cu}

The SIS multilayered structure concept offers perspectives of great RF performance enhancement but its implementation increases exponentially in complexity.
The required superconducting layers are much thinner than typical stand-alone superconducting films for cavity applications. So methods to achieve rapidly relaxed and high quality materials with nominal properties are required for optimized SIS performance. The introduction of dielectric layers complicates things further. 
Layers need to be well defined with sharp interfaces. Methods need to be optimized to deposit the superconductor on the insulator and the insulator on the superconductor maintaining the quality and properties of each layer and of the base substrate. 
Material choices and deposition methods have to be tailored to address material (inter-diffusion), environment (need of precursors, pressure), temperature (process versus substrate melting temperatures, inter-diffusion... ) compatibility.
For example, bulk-like properties are achieved at 600$^{\circ}$C both for NbTiN ($T_{c}$) and AlN (dielectric properties). However, at this temperature, Al diffuses into Nb and NbTiN, resulting in amorphous structures and diffuse interfaces when combining these materials in SIS structures \cite{Valente-Feliciano 2015}. Optimization for SRF performance requires then a compromise by lowering the deposition temperature to maintain the desired properties for both material layers and sharp interfaces.

The choice of techniques needs to remain mindful of the materials at play and the final SRF properties pursued. For example, ALD and CVD processes rely on the use of precursors, which are compounds often water based, or rich in H or C. This induces challenges in achieving the adequate purity and quality of the films while preserving the base materials. The final deposition scheme for optimal SIS multilayered structures may very well be a combination of deposition techniques depending on the chosen materials (energetic condensation for superconducting layers, ALD for insulators...). In this case, complex and integrated deposition systems need to be developed to allow in-situ processing and deposition of the functional SRF structures.

When transferring, deposition processes to 3D cavity structures, coating conditions can vary greatly at different locations (equator versus iris or beam tubes). In this context, the use of a sample host cavity, so-called "coupon cavity" that has detachable coupon samples at equivalent positions in elliptical cavities \cite{Saeki SRF2019} can prove to be a useful tool to test the thin film creation and assess the properties variation along the cavity profile. The coupon samples produced can then be evaluated by various characterization techniques.

Across all materials and deposition techniques, the final surface created needs to be stabilized in order to be impervious to all necessary procedures in the lifetime of an SRF cavity and preserve the RF layer responsible for the performance in operation. This can be achieved through controlled dry oxidation, deposition of a protective thin and RF transparent cap layer. The methods need once again be tailored to the material of interest.

\subsection{Improved cavity fabrication \& preparation techniques}
\subsubsection{Development of quality substrates}
    It is well recognized in thin film technology that the surface and material quality of the underlying substrate is of critical importance for the final properties of the deposited film \cite{Valente-Feliciano SRF 2011}.
    Recent studies have shed light on the extreme importance of the substrate quality to ensure the good SRF film properties. HIE-ISOLDE experience \cite{Hie-Isolde}, CERN campaign on 1.5 GHz spun and hydroformed \cite{CERN2006} and the most recent studies at CERN on electroformed \cite{Amador 2021}and bulk elliptical cavities have very clearly proven the superior performance of the superconductive films coated on a seamless substrate \cite{CERN 2021}. 
    Cavity structures cut in the bulk are a good research vehicle to establish the performance boundaries on films when deposited on ideal metallic substrates. They could also be considered for machines that require only a small number of cavities, such as for Hie-Isolde \cite{Hie-Isolde}. However, this path is not sustainable for accelerator machines requiring a large number of cavities. It is thus fundamental to continue the technology development of classical fabrication methods (spinning, hydroforming, electroforming) and investigate the potential of new production techniques such as electro-hydro forming and 3D additive manufacturing. 
    
    The substrate surface preparation is equally important, and new environmentally-friendly polishing techniques should be explored. Examples in this direction, even if some of them have been tested at the moment only on small surfaces, are mechanical polishing (Centrifugal Barrel Polishing, Diamond Turning), Laser Polishing \cite{Ries21}, Plasma Electrolytic Polishing. For any machine project requiring large numbers of cavities, it is imperative to find methods to reliably produce a substrate surface that is chemically stable. For instance, methods to efficiently and reproducibly passivate the surface of Cu cavities need developing as the native Cu oxide is highly unstable with  temperature and environmental conditions.
    
\subsubsection{Important considerations for substrates intended for alternative materials}
    
    The synthesis of A15 materials onto Cu substrates aims at improving the quality factor of coated Cu cavities by a factor of 10 at an operating temperature of 4.2 K.  Several challenges arise when using copper as the substrate for these compounds: 
    
    \begin{list}{-}{}
    	\item Thermal expansion coefficient difference between the substrate and the thin film may lead to a critical temperature depression due to residual stress build up 
  
    	\item     A high process temperature ($>300 ^{\circ}C$) is needed in order to form the proper A15 crystalline structure that is the only superconducting phase for those materials. The need of such high temperature may have a significant impact onto the design of the accelerating cavities that will have to be stiffened in view of compatibility with a tuning system during machine operation. 
 
    	\item     Interdiffusion of Cu into the A15 layer can spoil the superconducting layer and a strategy has to be developed to prevent such contamination.
    \end{list}

\subsubsection{Novel Cooling Techniques}
As they enable higher SRF cavity operation temperature ($\geq 4.2 K$), high-$T_{c}$ superconducting thin films will open the way for new conduction cooled accelerating structures using new cooling techniques (cavity wall with integrated liquid He cooling circuit or pulsating heat pipes, etc.) and cooling channels instead of helium tanks. Indeed, one of the major problems is the evacuation of the energy in-homogeneously deposited inside the cavity towards the cold source. Regardless of the superconducting film or structure used, improved heat transfer is essential. It is therefore necessary to provide innovative solutions that use existing and available technologies to ensure optimal heat transfer. Additive manufacturing of metals (Cu and Al elemental or alloys) then rises as an option for designing optimized thermal links and structures cooled by cryo-coolers. To fulfill this objective, several conditions need to be met. The materials chosen need to have optimized thermal conductivity, higher than for the SRF material. Exchange links and surfaces have to be optimized to increase heat transfers and minimize helium consumption. Other conditions are optimized mechanical properties for both material and cavity geometry along with compatibility with ultra-high vacuum and low surface roughness.

\subsection{Measurement platforms for relevant SRF film properties}

To investigate fundamental questions such as SRF loss mechanisms or material limitations, a form of RF measurement is a necessity in order to assess early SRF performance in the material process development, identify potential pitfalls and validate processes. One of the prominent challenges in novel SRF thin film  development is the parameter space to be explored. A development phase on samples with the implementation of specific material and RF performance diagnostics is then required before implementation on actual cavities. RF evaluation systems tailored to samples are of high interest and of high value, especially to perform AC measurements. Throughout the SRF community several systems exist, differing in geometry and sample size as well in the accessible parameter space of frequency, temperature and RF field range. For a review of available sample test cavities, the reader is referred to \cite{Goudket2017}. A versatile sample-test platform is the quadrupole resonator (QPR). It provides high resolution measurements ($<10n\Omega$) in a wide parameter space of temperature (1.5-20 K) and RF field up to 120 mT at three different frequencies (typically n x 433 MHz), covering the typical operational parameters of existing accelerators. Recent studies improved the resolution of such systems and solved a systematic operational problem, enabling QPR setups to reach their full potential for SRF R\&D \cite{Keckert2021}. 

QPR samples are relatively complex samples, much larger (about 75 mm in diameter, welded on a Nb tube) than the sample size ($< 10 mm$) typically employed to develop new films and deposition techniques. In order to optimize resources and time at the sample stage, it is important to find DC characterization techniques that can give indication of the sought-after RF properties ($j_{c}$,$H_{c}$,…) and to establish the potential of the techniques early in the development process.
This is specifically important for the assessment of complex materials and SIS structures. 

A common characterization method for superconducting small flat samples is DC magnetometry. However in most setups, the sample is immersed in a DC field with some orientation and edge effects due to the demagnetization parameter. In order to overcome this issue, a local magnetometer based on the $3^{rd}$ harmonic method was developed at Saclay \cite{Antoine 2010, Aburas SRF2017, Antoine 2019}, where the parallel magnetic field is applied with a coil which size is much smaller than the sample size. As the field decays quickly away from the coil, the sample can be considered as an infinite plane and no demagnetization effects occur. The applied parallel magnetic field also mimics the magnetic field configuration at the equator of SRF cavities. The good results achieved at CEA Saclay stimulated collaborations with other institutions to develop similar platforms (Kyoto University, KEK, JLab) and extend the capability. 

It is important to investigate SRF and superconducting properties beyond surface resistance, to include structure/property relations that impact electrodynamic and thermal performance at the extreme limits present in SRF cavities.  Local probes that produce quantitative measurements of critical RF properties of coatings and films are of great utility and interest.  Examples include cryogenic scanning electron and force microscopies to reveal the presence of surface hydrides, scanning tunneling microscopy to identify states in the gap and defects/adatoms that create zero-bias conductance peaks, and scanning microwave microscopy to measure local sources of nonlinear response and correlate them with specific defects and near-surface features. 

Tunneling spectroscopy (TS) is a 50 years old DC technique that enable the mapping of the fundamental surface superconducting properties at cryogenic temperature and under external magnetic field. From these measurments, parameters extracted and fits ($T_{c}$, superconducting gap $\Delta$) provide in-depth understanding of microscopic dissipation phenomena. TS has been successfully applied to numerous SRF cavity cut-outs or samples treated in a similar manner as cavities and data analysis has shown predictive capability of SRF performance ($Q_{0}$, $E_{max}$) on Nb and Nb$_{3}$Sn \cite{Groll2022, Becker2015}. 

As mentioned before,superconducting and RF properties are inherent to the material.  Thus characterization cannot be only limited to superconducting and RF properties, it needs to include surface and material. In depth material and surface studies require multiple characterization techniques (STEM, TEM, FIB, EDS, EBSD, XRD, XPS, SIMS...). These are often very specialized and require expensive major instrumentation and operator expertise.

\subsection{Cryomodule design optimization}

Once thin-film coated cavities attain performance specifications for accelerator application, the integration of these cavities into an operational cryomodule may encounter several technical challenges depending on the mechanical and material properties of thin-film materials and the substrates. Accordingly, it may demand modification in the design of current cryomodules and their operations to preserve the performance of such cavities. Many technical limitations may not be recognized until we progress toward the cryomodule phase. Still, the R\&D efforts should be started early to identify them in advance and develop mitigation strategies. Recent experience with Nb$_{3}$Sn-coated cavities indicates that they are mechanically vulnerable because of the brittle nature of the material. So that the tuning scheme in the cryomodule should be adjusted or modified to preserve the coating, most thin-film coated cavities require slow cool-down for optimal performance. They should be accommodated during cryomodule cool down. Focused scientific and technical efforts should be started sooner instead of later to identify and mitigate such issues to optimize cryomodule design to integrate thin-film coated cavities.
On the other hand, the strategy of decoupling the SRF surface from the mechanical support structure can open the way to overal cryomodule design simplification.

\subsection{Improvement of accelerator ancillaries with advanced deposition techniques (HiPIMS Cu coated bellows, power couplers…)}
 Novel thin film deposition techniques as described above also have a strong potential for the development of accelerator ancillary components. For example, many SRF systems design call for the use of Cu plated bellows and power couplers. Traditionally these components are fabricated with Cu electroplating on stainless steel. This technique produces layers with a large variation of thickness across the part profile, typically 200 \%. The detailed process is often considered a trade secret and the performance is strongly tied to the operator. This generally has led to the need of re-developing the process time and time again over the years for the need of various accelerator projects.  Recent developments in HiPIMS deposition of Cu on stainless steel, specifically with added Kick technology \cite{Starfire} are showing improved thickness uniformity and adhesion to the substrate without the need for a seed  or interlayer (typically Ni, a magnetic material). It is expected that the technique development with well defined parameters is highly reproducible and can easily be transferred from one provider to another.  
 
\subsection{Preparation for industrialization }

Large scale projects require high reproducibility in performance. Technology transfer to industry also requires robust processes. With well defined process steps in highly controlled conditions, thin film technology offers strong levels of reliabiltiy and reproducibility. Thus, early development of automation is key for results reliability with SRF thin film technology deployment:   
As a high quality surface preparation is key for the success of all subsequent process steps, as frequently observed in industrial coatings, automation of surface treatment processes and substrate handling can remove factors of variability.

The path to consistently high gradients, small form factors and, ultimately, industrialization of the whole concept will require a high grade of automation of the coating process.

\section{R\&D Roadmap for Next Generation SRF Thin Film Cavities}

\begin{figure}
 	\begin{center}
 		\includegraphics[width=1\hsize]{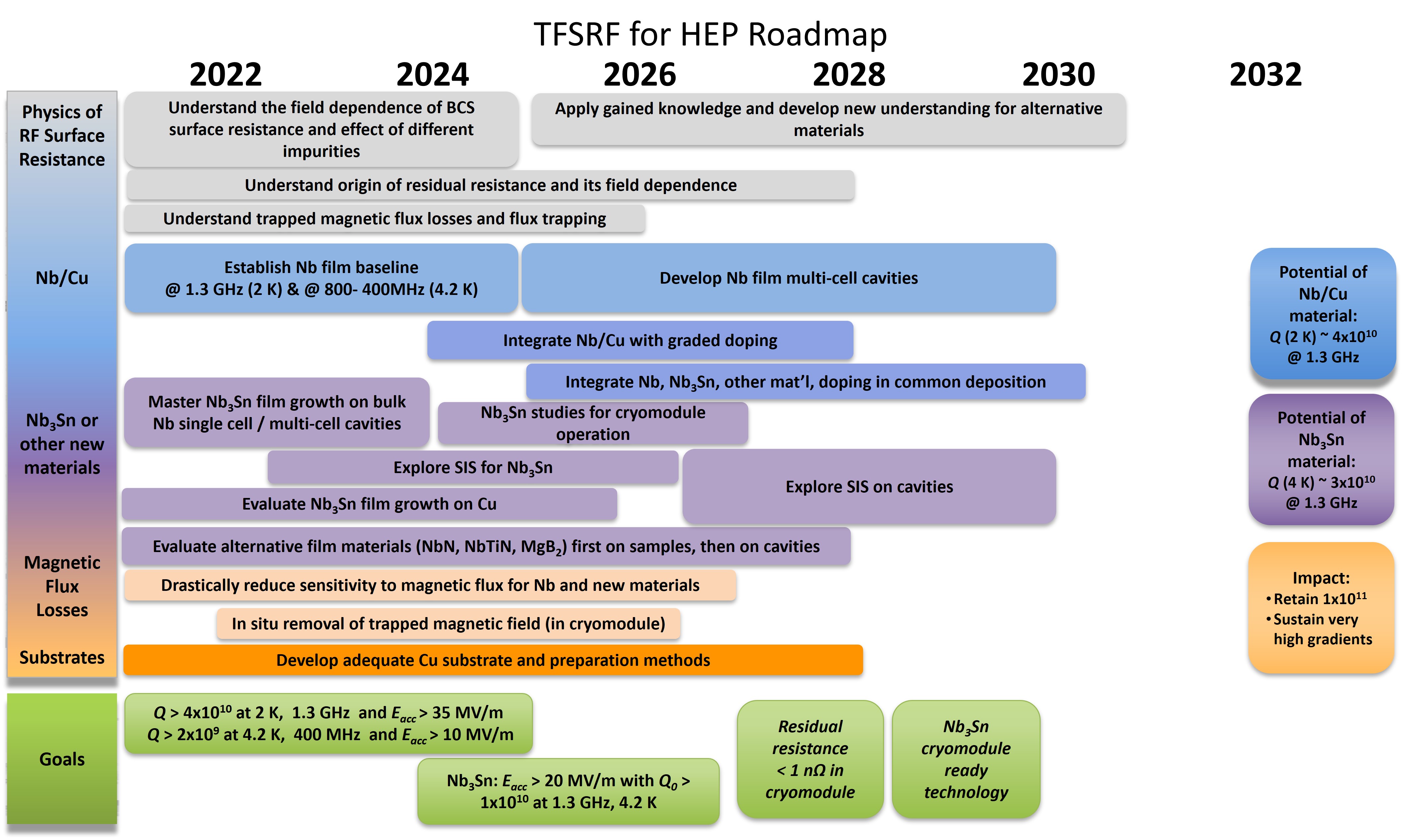}
 	\end{center}
 	\caption{HEP Roadmap for STF Thin Film Technology development, adapted from 2017 US SRF Roadmap \cite{SRFRoadmap 2017}}
 	\label{fig:TFSRFRoadmap}
 \end{figure}

Efforts in the United States for SRF thin films research and development should pursue the three main research thrusts, in synergy with other regions, Europe and Asia with the roadmap and milestones described therafter (See Fig.~\ref{fig:TFSRFRoadmap}).  This roadmap is coherent with the European Strategy for particle physics document published January 2022 \cite{EU Strategy 2022}.

\subsection{Nb/Cu Technology}  

The demonstration of Nb/Cu performance is paramount since it remains the simplest system after bulk Nb and other materials/structures development imply increasing difficulty and complexity.  
The milestones accomplished in the next five years should: 
\begin{list}{-}
	\item Establish the Nb/Cu baseline at 1.3 GH with demonstration of quality factors at 2 K of 2x$10^{10}$ at 25 MV/m and then at 35 MV/m.
	
	\item Scale-up Nb/Cu technology to lower frequency cavities (800 - 400 MHz) and establish baseline 
\end{list}
The objectives for the following five years should be to:
\begin{list}{-}
	\item Develop Nb/Cu film deposition on multi-cell accelerator cavities  

	\item Integrate graded doping/alloying on Nb/Cu film cavities
\end{list}

\subsection{Alternate superconductors}  

The degree of advancement in research for alternative materials to Nb varies drastically depending on the material at play.
The following milestones only concern the Nb$_{3}$Sn development by alternative techniques and other A15 compounds, MgB$_{2}$. The roadmap for Nb$_{3}$Sn by the now standard method of Sn diffusion is captured in the dedicated white paper \cite{Nb3Sn-SnowmassWP}.

In order to fully explore the potential of alternative materials for 4.2 K operational cost saving, it is essential to pursue their development specifically with novel coating methods, compatible with low temperature substrate materials. 

The development of an alternate material, Nb$_{3}$Sn or other, for SRF applications requires in the next decade:
\begin{list}{-}{ }
	\item The development of novel coating techniques for Nb$_{3}$Sn, other selected A15 compounds (Nb$_{3}$Al, V$_{3}$Si, Mo$_{3}$Re) and MgB$_{2}$.
	\item The parameter optimization for the technique(s) of choice
	\item The transfer of the deposition technique(s) to 3D coating of 1.3 GHz Nb or Cu single-cell cavity
	\item The demonstration of performance at 4.2 K and 2 K similar to Nb$_{3}$Sn cavities produced by Sn vapor diffusion.   
	\item  The identification of issues and elaboration of solutions
	\item The deployment of  cavity deposition to multi-cell cavities 
	
\end{list}

\subsection{Multilayered SIS structures}  
The envisionned exponential benefit in SRF performance from multilayered SIS structures is met by exponential complexity in developments, requiring a coordinated long-range research plan.

Explore SIS structures potential with Nb based B1 compounds and Nb$_{3}$Sn .   

\begin{list}{-}{ }
	\item Develop deposition schemes in function of chosen materials, keeping in mind compatibilities of materials and deposition methods.
	\item Optimize deposition parameters to maximize interface refinements and combined properties 
	\item Evaluate the RF potential of such structures on large samples with RF measurements by QPR and DC characterization (magnetometry, ...).
	\item Transfer to 3D geometries
	\item Demonstrate the elaboration and performance of simple SIS structures on Nb cavities.
	\item Identify potential issues and develop solutions
	\item Demonstrate maximum field enhancement with the fore-mentioned SIS structures first on Nb bulk cavities, then on Nb/Cu cavities.
	\item For industrial applications, demonstrate SRF performance for cavities operated in cryo-coolers.
\end{list}

\subsection{Supporting Developments}
 As substrates are critical, simultaneously with SRF films and structure development, research needs to be pursued to produce cavity support structures which are mechanically and chemically stable, offering the best possible template for the SRF thin film structures.
 
 As challenges are identified, they need to be fully addressed during the R\&D. 
 Novel tuning methods are required to preserve delicate materials such as the A15 compounds. The use of MgB$_{2}$ will most likely require the development of capping layers to preserve the material.

 %%%%%%%%%%%%%%%%%%%%%%%%%%%%%%%%%%%%%%%%%%%%%%%%%%%%%%%%%%%%%%%%%%%%%%%%%
 %%%%%%%%%%%%%%%%%%%%%%%%%%%%%%%%%%%%%%%%%%%%%%%%%%%%%%%%%%%%%%%%%%%%%%%%%%%

\section{Synergies and Community Needs}

Superconducting RF is a key technology for future HEP and NP accelerators, now relying on advanced surfaces beyond bulk Nb for a leap in performance and efficiency. With dedicated and sustained investment in SRF thin film R\&D, next-generation thin-film based cavities will become a reality with high performance and efficiency,  facilitating energy sustainable science while enabling higher luminosity, and higher energy.  
\paragraph{Collaboration \& synergies}
As discussed in the previous section, the field of needed developments for SRF thin film technologies to reach their full potential is vast and diverse.  
Close collaborations within the United States and with institutions worldwide and synergies between their dedicated R\&D programs are of decisive importance. 
There exist already well-established and fruitful international R\&D collaborations involving JLab, FNAL, SLAC, ODU, UMD, CERN, INFN-LNL, CEA Saclay, STFC, HZB, DESY, KEK, and other institutions. The number of contributing institutions to the field is growing, bringing in additional expertise. This has led to the establishment of a workshop series dedicated to SRF thin films with its first edition organized in 2005 as a spin-off of the bi-annual SRF Conference. Since then, the \textit{"International Workshop on Thin Films and New Ideas for SRF"},organized typically bi-annually, has seen its participation grow over the years. It aims at providing a forum for new initiatives in innovative thin films and related technology to advance future generations of SRF accelerators, infused by expertise of specialists from related disciplines. The TTC collaboration has also mandated a dedicated SRF Thin Film working group to facilitate dialogue and problem solving across the community.

\paragraph{Scientific tools} 

As mentioned before, SRF is a multidisciplinary field which requires correlation between surface, material properties and RF performance. Access to major material research and characterization instruments along with the associated expertise is mandatory for fundamental and detailed studies and should be facilitated through funding and strong collaborations with institutions both in the US and internationally that already own the capabilities.  

Sustained progress in the development of engineered functional superconducting thin films and structures with enhanced RF performance (Nb/Cu, Nb$_{3}$Sn, other A15, other superconductors (B1 compounds, MgB$_{2}$...), multi-layers, integrated doping or alloying, peak fields...) requires a high throughput of samples and cavities testing. Thus it demands availabilty of multiple RF test platforms such as the QPR, …).  

\paragraph{Dedicated and sustained funding} 

In order for SRF thing film technologies to mature and enable efficient and highly performing SRF systems for future accelerators, dedicated funding streams need to support both fundamental R\&D and engineering developments. Significant levels of funding are typically tied to specific machine projects which in turn are assuming mature, industry and large-scale ready technologies. 

The promised strides in the developments described here will only come to realization with adequate labor resources, state-of-the-art production and test facilities (cleanrooms, chemistry setups, RF testing, major characterization instruments), including existing facility upgrades and new facilities in step with technology advances.

\paragraph{Development of tomorrow SRF workforce}  
The development of SRF thin film technologies offers extensive training opportunities to shape the next generation of SRF experts. The future of HEP in the United States and beyond relies on the development of young scientists and engineers who will take on the implementation of future accelerator facilities and industrialization of SRF technology.  Among them will rise future leaders that will think, conceive, develop and build the accelerator machines that will enable over the next decades future research programs for HEP and other sciences beyond. 

\paragraph{Fostering industrial partners in US}  
Industrialization of SRF systems has been up to now limited. The past decade has seen US cavity industrial providers walking away from this activity due to the necessary infrastructure investments too significant relative to the size of the available market. Most of the SRF cavities for US DOE large projects such as LCLS-II, SNS PPU, are now produced outside the United States. On the other hand, small US companies have contributing profusely to SRF related R\&D projects funded through the DOE SBIR program. One could foresee with the application of SRF thin films to accelerators for societal needs that the US and international markets could grow significantly. It is thus important to engage industrial partners early in the developments linked to SRF thin film technology.

\section{Conclusion and Recommendations}

SRF thin film technology based on advanced coating techniques offers many opportunities to fully engineer functional SRF surfaces with the deliberate creation of the most favorable interface or interlayer, tailoring of the most favorable film(s) structure, properties enhancement with doping/alloying, and control over the final RF surface with dry oxidation or cap layer protection. Highest performing Nb films, alternative material films and SIS multilayered structures open the possibility of enhanced performance beyond bulk Nb and major system simplifications.  Such developments would be transformative not only for future high energy physics machines but will also bring forth the opportunity to upgrade existing machines to higher performance in achievable energies and cryogenic \& power consumption, within the same footprint. 

The already well-established and fruitful international R\&D collaborations involving US institutions and their international counterparts should be fully supported and expanded in the following areas of R\&D:  

\begin{list}{-}{ }
	\item Theoretical and material studies to gain in-depth understanding of the fundamental limitations of thin film superconductors under radio-frequency fields 
	\item Advanced coating technology via energetic condensation (electron cyclotron resonance (ECR), HiPIMS, kick positive pulse…), via Atomic Layer Deposition (ALD) and other hybrid techniques depending on the material requirements, Nb/Cu, alternative materials: Nb$_{3}$Sn, V$_{3}$Si, NbTiN … 
	\item Development of superconductor-insulator-superconductor (SIS) nanometric layers to further enhance the performance of bulk Nb and Nb/Cu cavities
	\item Improved cavity fabrication \& preparation techniques (electroforming, spinning, hydroforming, electro-hydro forming, 3D additive manufacturing, environmentally friendly electropolishing, nano-polishing, plasma etching …) 
	\item Cryomodule design optimization tailored to the SRF material of choice
	\item Improvement of accelerator ancillaries with advanced deposition techniques (HiPIMS Cu coated bellows, power couplers…) 
\end{list}

The progress achieved thus far in SRF thin film research constitutes a reasonable stepping stone to truly “engineered SRF surfaces”, with all the benefits of high $Q$, high field, low cost, high reliability systems.
As resources drive the timeline,continued investment is required so this fundamental R\&D can mature and become project ready for the next generation accelerators and we ask for the Snowmass community affirmative support in this endeavor.

%%%%%%%%%%%%%%%%%%%%%%%%%%%%%%%%%%%%%%%%%%

%  If you would like to use BibTEX for the bibliography, please feel free to do so.  It is not required.

%  To use BibTeX,

%    1.  uncomment the following two lines, 
%    2.  comment out everything below from  \begin{thebibliography}{99}   to \end{thebibliography).
%    3.  create the file  myreferences.bib, and process this file in the usual way

%\bibliographystyle{JHEP}
%\bibliography{myreferences}  % file myreferences.bib

%%%%%%%%%%%%%%%%%%%%%%%%%%%%%%%%%%%%%%%%%

\end{document}